\documentclass[twocolumn,amsmath,amssymb]{revtex4-2}

\usepackage{bm} 
\usepackage{braket} 
\usepackage{lipsum} 
\usepackage{mathdots} 
\usepackage[version=3]{mhchem} 
\usepackage{xspace} 
\usepackage{url}

\usepackage[pdftex]{hyperref} 
\usepackage[pdftex]{graphicx}

\renewcommand{\_}[1]  {_\textrm{#1}}

\newcommand{\CSS}     {\ce{Co3Sn2S2}\xspace}

\begin{document}

\title{
    Effective model analysis of intrinsic spin Hall effect with magnetism 
in stacked-kagome Weyl semimetal \CSS
}

\author{Akihiro Ozawa$^1$}\thanks{Present address : Institute for Solid State Physics, The University of Tokyo, Kashiwa 277-8581, Japan}\thanks{akihiroozawa@issp.u-tokyo.ac.jp}
\author{Koji Kobayashi$^2$}\thanks{Present address: Physics Division, Sophia University, Chiyoda-ku, Tokyo 102-8554, Japan}\thanks{k-koji@sophia.ac.jp}
\author{Kentaro Nomura$^2$}\thanks{nomura.kentaro@phys.kyushu-u.ac.jp}

\affiliation{$^1$Institute for Materials Research, Tohoku University, Sendai 980-8577, Japan}
\affiliation{$^2$Department of Physics, Kyushu University, Fukuoka 819-0395, Japan}

\begin{abstract}
~We theoretically study the spin Hall effect in a simple tight-binding model of stacked-kagome Weyl semimetal \ce{Co3Sn2S2} with ferromagnetic ordering.
We focus on the two types of the spin Hall current: one flowing in the in-plane direction with respect to the kagome lattice~(in-plane spin Hall current), and one flowing in the stacking direction~(out-of-plane spin Hall current).
We show the spin Hall conductivities for those spin currents drastically change depending on the direction of the magnetic moment.
Especially, the out-of-plane spin Hall current may induce surface spin accumulations, which are useful for the perpendicular magnetization switching via spin-orbit torque. 
\end{abstract}

\maketitle

\section{Introduction}
 The generation
and control of spin current, namely
the flow of spin angular momentum, are important objectives in spintronics.
Spin Hall effect~(SHE)
\cite{Dyakonov1971,Kato2004,Sinova2015}
is one of the most fundamental phenomena for generating spin current;
the spin current is driven transversely to an applied electric field.
Spin-orbit coupling~(SOC) plays a significant role in obtaining the spin-dependent electron motion and thus the SHE.
The SHE-based highly-efficient manipulation of magnetization,
such as spin-orbit torque \cite{chernyshov2009,miron2010,miron2011,liu2012current,liu2012spin,fukami2016},
has recently been studied.
Conventionally, non-magnetic materials with a strong SOC, such as Pt~\cite{Guo2008} and Ta~\cite{liu2012spin}, had been examined as a spin Hall current generator.
Meanwhile, in such a system, the direction of the accumulated spin 
at the interface is constrained to be perpendicular to both the direction of the flow of spin current and the applied electric field~\cite{Dyakonov1971,Sinova2015}.

In addition to non-magnetic materials, recent studies explore the possibility of magnetic materials as spin Hall systems~\cite{Miao2013,Taniguchi2015,Seki2015,Kimata2019,Wang2021,Huyen2021,Hu2022}.
Particularly, some magnetic systems exhibit a peculiar SHE where the direction of the accumulated spin is parallel to that of flow of spin current, 
depending on the magnetic configurations, such as antiferromagnetic ordering~\cite{nan2020,kondou2021,Hu2022}.
Finding spin Hall systems without restrictions on the direction of induced spin accumulation may help us design functional spintronic devices.

Recently, from viewpoint of spintronic functionalities, Weyl semimetals with magnetism, magnetic Weyl semimetal have been studied
~\cite{araki2016,kuroda2017,araki2018,sakai2018,he2022,maekawa2023}. 
Originating from the Berry curvature generated by the band crossing points~(Weyl points) ~\cite{Wan2011,Burkov2011},
distinctive electromagnetic responses,
such as the intrinsic anomalous Hall effect~(AHE)~\cite{Burkov2011}, occur.
ABC-stacked kagome-lattice ferromagnet \CSS is a promising candidate for magnetic Weyl semimetal~\cite{Liu2018,Xu2018,Wang2018,liu2019arpes}.
The giant AHE arises because the Weyl points are very close to the Fermi level.
Additionally, the giant anomalous Hall angle is realized due to the small longitudinal conductivity, 
i.e., the small Fermi surfaces~\cite{Liu2018}.
Other characteristic responses, for example, the anomalous Nernst effect~\cite{guin2019} and magneto-optical Kerr effect~\cite{okamura2020}
are also studied.
According to the above features, we expect \CSS might be useful for the efficient and functional manipulations of the spin current for the following reasons. 
First, the fully-polarized current is realized because 
of the half-metalicity~\cite{Liu2018,Lin2019}.
Thus the giant anomalous Hall angle implies highly-efficient SHE.
Second, owing to SOC, the band topology and AHE depend on the magnetic configurations~\cite{Ghimire2019,zhang2021unusual,watanabe2022}.
Similarly, the characteristic relations between the SHE and the direction of the magnetic moments are expected.

~In this paper, 
we theoretically investigate the SHE and AHE in an effective tight-binding model ~\cite{Ozawa2019} of magnetic Weyl semimetal \CSS with ferromagnetic ordering.
First, we review that this model can describe the Dirac semimetal state and the SHE in a non-magnetic state.
Then we show that spin Hall conductivities~(SHCs) drastically change depending on the direction of the magnetic moment.
Especially, the out-of-plane SHE is enhanced by tilting the magnetic moment from  the $z$ axis, in the presence of a certain SOC.
Lastly, the possibility of this system as a spin-current generator for out-of-plane magnetization switching is discussed.

\section{Tight-binding Model}
\begin{figure}[t]
\includegraphics[width=1.0\hsize]{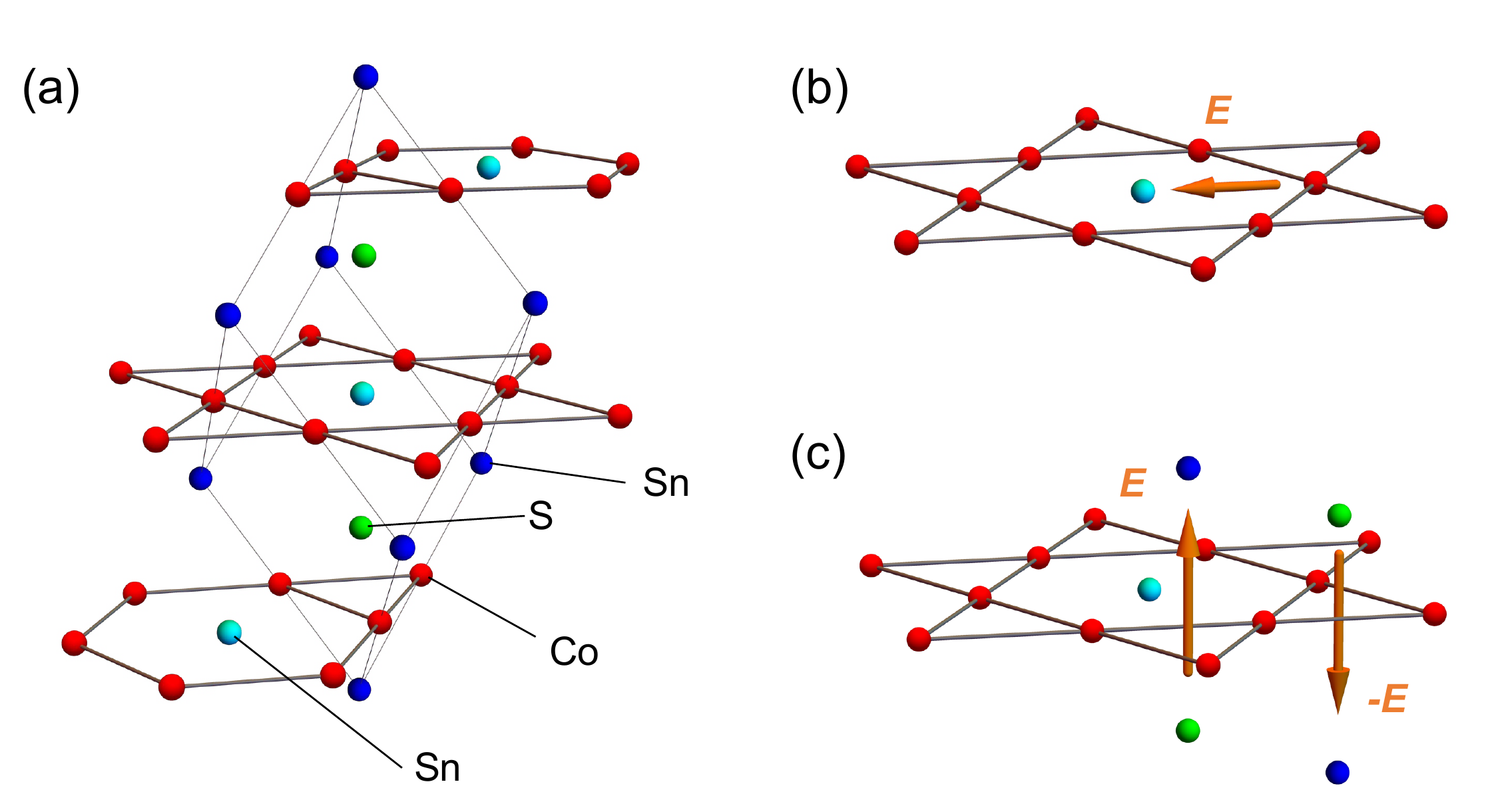}
\caption{
(Color online) 
(a)~Crystal structure of \CSS. 
(b)~Electric field generated by Co nucleus~(red) and Sn nucleus~(cyan) at the center of the hexagon in the kagome layer, arising intra-layer-kagome type SOC.
(c)~Electric field generated by Sn nucleus~(blue) and S nucleus~(green) in between the kagome layers,
arising the staggered-Rashba type SOC.
} 
\label{fig:SOC}
\end{figure}

First we explain the minimal tight-binding model of magnetic Weyl semimetal \CSS introduced in our previous study Ref.~\onlinecite{Ozawa2019}.
This model describes the six pairs of the Weyl points by using the $d$ orbital of Co and the $p$ orbital of Sn.
Figure \ref{fig:SOC} shows the crystal structure of \CSS. 
The kagome layers consist of Co, being responsible for magnetism.
Sn atoms are located at the center of these hexagons of the kagome lattice.
Triangular lattice layers formed by one Sn and two S are sandwiched by the kagome layers.
In our effective model, one $d$ orbital from Co forming the kagome layers and one $p$ orbital from interlayer Sn are extracted.
We neglect the rest of the orbitals for simplicity.
Therefore the unit cell consists of the (3+1) sublattices.
The primitive translational vectors are defined as 
$\bm{a}_1=(\frac{a}{\sqrt{3}},0,\frac{c}{3})$, 
$\bm{a}_2=(-\frac{a}{2\sqrt{3}},\frac{a}{2},\frac{c}{3})$, 
$\bm{a}_3=(-\frac{a}{2\sqrt{3}},-\frac{a}{2},\frac{c}{3})$.
Here, $a$ and $c$ are lattice parameters of the conventional unit cell, given as $a=5.37$\AA\xspace and $c=13.18$\AA~\cite{Liu2018}, respectively.
The total Hamiltonian of this model is given by,
\begin{align}
 H_{\text{0}}=H\_{hop}+H\_{soc}+H\_{exc}.
\label{eqn:hamiltonian}
\end{align}
Here, 
$H\_{hop}$ is the spin-indenpendent hopping term,
$H\_{soc}$ is the SOC term,
and $H\_{exc}$ is the exchange coupling term between electron spins and the magnetic moment.
$H\_{hop}$ is given by,
\begin{align}
 H\_{hop}=- \sum_{ijs}t_{ij} d^{\dagger}_{is}d_{js}+t\_{dp}\sum_{\langle ij \rangle s}(d^{\dagger}_{is}p_{js}+p^{\dagger}_{is}d_{js})   \nonumber\\
+\epsilon\_{p}\sum_{is} p^{\dagger}_{is}p_{is},
\end{align} 
where $d_{is}$ and $p_{is}$ are the annihilation operators of the Co-$d$ and Sn-$p$ orbitals, respectively.
The index $s$ is for the spin, and $i$ for the site.
$t_{ij}$ includes the first and second-nearest neighbor hopping, $t_1$ and $t_2$, in the kagome layer, inter-kagome layer hopping $t_z$. 
The summation for $\langle ij \rangle$ is taken over the nearest neighbor hopping between the Co and Sn sites.
 $t\_{dp}$ represents the $dp$ hybridization between the Co-$d$ and Sn-$p$ orbitals.
The lattice vectors for the intra-layer nearest-neighboring are calculated by 
${\bm b}_{AB}=({\bm a}_2-{\bm a}_1)/2$,
${\bm b}_{BC}=({\bm a}_3-{\bm a}_2)/2$, and 
${\bm b}_{CA}=({\bm a}_1-{\bm a}_3)/2$.
In the same manner, the lattice vectors for intra-layer second-nearest-neighboring are calculated by 
${\bm d}_{AB}=({\bm b}_3-{\bm b}_2)/2$,
${\bm d}_{BC}=({\bm b}_3-{\bm b}_2)/2$, and 
${\bm d}_{CA}=({\bm b}_1-{\bm b}_3)/2$.
The lattice vectors for the inter-layer nearest-neighboring are calculated by 
${\bm c}_{AB}=({\bm a}_1-{\bm a}_2)/2$,
${\bm c}_{BC}=({\bm a}_2-{\bm a}_3)/2$, and 
${\bm c}_{CA}=({\bm a}_3-{\bm a}_1)/2$.
Here, A,B and C are the sublattice indices \cite{Ozawa2019}.
\begin{figure*}[t]
\includegraphics[width=1.0\hsize]{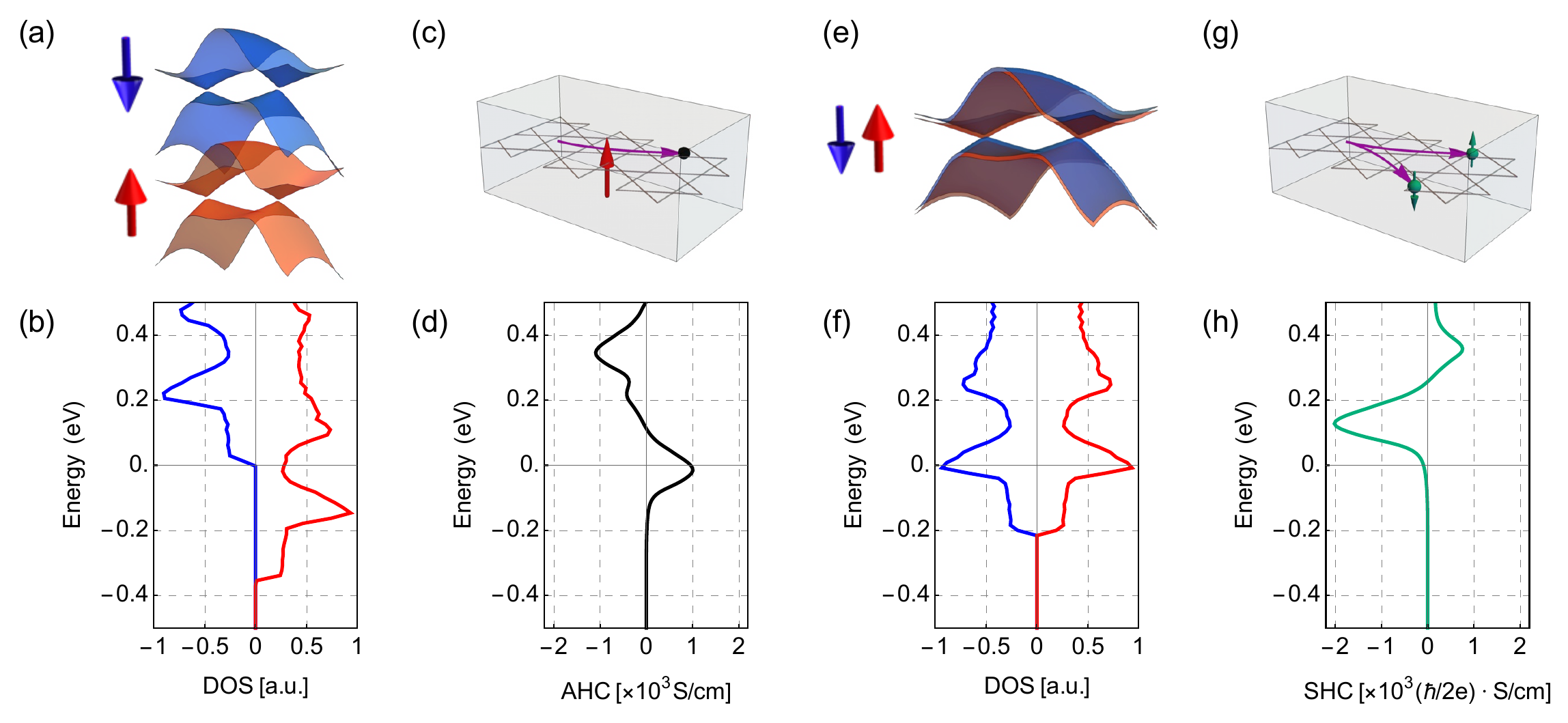}
\caption{(Color online)~
For magnetic Weyl semimetal state in this model,
(a)~schematic figure of band structure,
(b)~density of states,
(c)~schematic figure of anomalous Hall effect, and
(d)~anomalous Hall conductivity.
For Dirac semimetal state,
(e)~schematic figure of band structure,
(f)~density of states,
(g)~schematic figure of spin Hall effect, and
(h)~spin Hall conductivity.
Fermi level is calculated with a constant electron number $n\_{e}=3$ per unit cell.
}
\label{fig:weyl_dirac}
\end{figure*}
The SOC term is given by $H\_{soc}=H\_{KM} +H\_{sR}$, where
\begin{align} 
 H\_{KM}=-\text{i}\it{t}\_{KM} \sum_{\braket{\!\braket{ij}\!} ss'} \nu_{ij} \cdot d^{\dagger}_{is} { \sigma}^z_{ss'} d_{js'},
\label{eqn:KM}\\
 H\_{sR}=-\text{i}\it{t}\_{sR} \sum_{\langle ij \rangle ss'} {\bm \lambda}_{ij} \cdot d^{\dagger}_{is} {\bm \sigma}_{ss'} d_{js'}.
\label{eqn:stgRashba}
\end{align}
$H_{\rm KM}$ describes the intra-kagome-layer Kane-Mele type SOC \cite{Kane2005,Guo2009} with strength $t\_{KM}$.
${\bm\sigma}$ is the vector of Pauli matrices,
corresponding to the electron spin.
The sign is $\nu_{ij}=+1(-1)$
when the electron hops counterclockwise~(clockwise) to reach the second-nearest-neighbor site on the kagome plane.
The summation $\braket{\!\braket{ij}\!}$ is taken for intralayer second-nearest-neighbor sites.
As shown in Fig.~\ref{fig:SOC}(b), this SOC originates from the potential of Sn at the center of hexagons of the kagome lattice.
Then we explain the staggered-Rashba type SOC $H_{\rm sR}$ as introduced in Ref.~\onlinecite{Chen2014}.
In \CSS, this SOC originates from the local inversion symmetry breaking of the Sn and S sites, as discussed below.
As shown in Fig.~\ref{fig:SOC}(c),
when we focus on one of the triangular plaquettes of the kagome lattice,
Sn is located on the top while S is located on the bottom.
Its nearest neighboring plaquettes have the reversed configuration.
Therefore, a perpendicular electric field penetrates each triangular plaquette in a staggered pattern.
$\bm{\lambda}_{ij}$ denotes the effective magnetic field vector from the electric field when the electron hops from the site $j$ to $i$.
${\bm \lambda}_{AB}=(\frac{1}{2},\frac{\sqrt{3}}{2},0)$, 
${\bm \lambda}_{BC}=(-1,0,0)$, 
${\bm \lambda}_{CA}=(\frac{1}{2},-\frac{\sqrt{3}}{2},0)$.
The summation $\langle ij \rangle $ is taken for intralayer nearest-neighbor sites.
This staggered-Rashba type SOC has a significant role in obtaining the finite out-of-plane AHE and SHE, as discussed later.

The exchange coupling between the itinerant electron's spins and magnetic moments on the kagome lattice is given by,
\begin{align}
 H\_{exc} = -J \sum_{iss'} {\bm m}_{i} \cdot (d^{\dagger}_{is} {\bm \sigma}_{ss'} d_{is'}
               + p^{\dagger}_{is} {\bm \sigma}_{ss'} p_{is'}).
 \label{eqn:exc}
\end{align}
$J$ is the exchange coupling constant and ${\bm m}_i$ is the vector of magnetic moment.
We note that ${\bm m}_i$ should be calculated self-consistently with the Coulomb interaction~\cite{Ozawa2022}.
Meanwhile, we 
We set the same value of exchange coupling on Co and Sn site for simplicity. 
In the following, we set $t_1$ as a unit of energy, 
$t_2=0.6t_1$, 
$t\_{dp}=1.8t_1$, 
$t_{z}=-1.0t_1$, 
$\epsilon\_{p}=-7.2t_1$,
$t\_{KM}=-0.1t_1$,
$t\_{sR}=0.1t_1$,  and   
$J=1.2t_1$.
These parameters are also chosen to fit the band structure to the result obtained by the first-principles calculations~\cite{Liu2018,Yanagi2021}. 
The Bloch Hamiltonian matrix $ \mathcal{H}(\bm{k})$ can be written in the form,
$H
=\sum_{\bm{k}s} C^{\dagger}_{\bm{k}s} \mathcal{H}(\bm{k}) C_{\bm{k}s},$
where
$C_{\bm{k}s}^{\dag}
=(d^{\dag}_{\bm{k}As}, d^{\dag}_{\bm{k}Bs}, d^{\dag}_{\bm{k}Cs}, p^\dag_{\bm{k}s})$.
In the following, we study the SHC and AHC based on this tight-binding model.

\section{Ferromagnetic Weyl semimetal state and non-magnetic Dirac state}

To keep this paper self-contained, 
here we review the ferromagnetic Weyl state and the paramagnetic Dirac state in the model \cite{Lau2022}.
In Fig.~\ref{fig:weyl_dirac}(a), 
the energy spectrum of the spin-splitting Weyl state is schematically shown.
Figure~\ref{fig:weyl_dirac}(b) is the density of states~(DOS) as a function of the energy, computed with ${\bm m}=(0,0,1)$.
We use $t_1=0.15$~eV as a unit of energy. 
$E=0$~eV is set as $E\_{F}$ obtained by the electron number per unit cell $n\_{e}=3$, corresponding to $E\_{F}$ of non-doped \CSS~\cite{Ozawa2019,Ozawa2022}.
Here we neglect staggered-Rashba type SOC because that does not affect the discussion in this section qualitatively. 
The DOS shows a local minimum 
near $E\_{F}$, corresponding to the energy of the Weyl points.
In the Weyl semimetal state, the AHE occurs, as schematically shown in Fig.~\ref{fig:weyl_dirac}(c).
Figure~\ref{fig:weyl_dirac}(d) shows $\sigma^{\rm AHE}_{yx}$ as a function of energy.
Here, the AHC is calculated by the Kubo formula \cite{Nagaosa2010} as

\begin{align}
 \sigma^{\rm AHE}_{yx}=
  & e^2 \hbar \sum_{n\neq m} {\rm Im}\int_{\rm BZ}\frac{d^3k}{(2\pi)^3} 
    \frac{f(E_{n{\bm k}})-f(E_{m {\bm k}})}{(E_{n{\bm k}}-E_{m{\bm k}})^2} \nonumber\\
  &\times
   \Braket{n\bm{k}| \hat{v}_y|m\bm{k}}\Braket{m \bm{k}|\hat{v}_x|n\bm{k}}
 \label{eqn:kubo_ahc_ip}.
\end{align}
$\hat{v}_{i}= \frac{1}{\hbar} \frac{ \partial H({\bm k})}{\partial k_i}$ $(i=x,y)$
is the velocity operator.
$f$ is the Fermi-Dirac distribution function
with $k\_B T = 0.01 t_1$.
The AHC is maximized near the $E\_{F}$, originating from the Berry curvature generated by the Weyl points.
As we discussed in our previous study~\cite{Ozawa2019}, the value $\sim$1000 $[\rm{S/cm}]$ is very close to the result obtained by the first-principles calculations and experiment~\cite{Liu2018}.
Therefore, our model describes the ferromagnetic Weyl state with the small DOS and large AHC in \CSS.

\begin{figure*}[t]
\includegraphics[width=0.9\hsize]{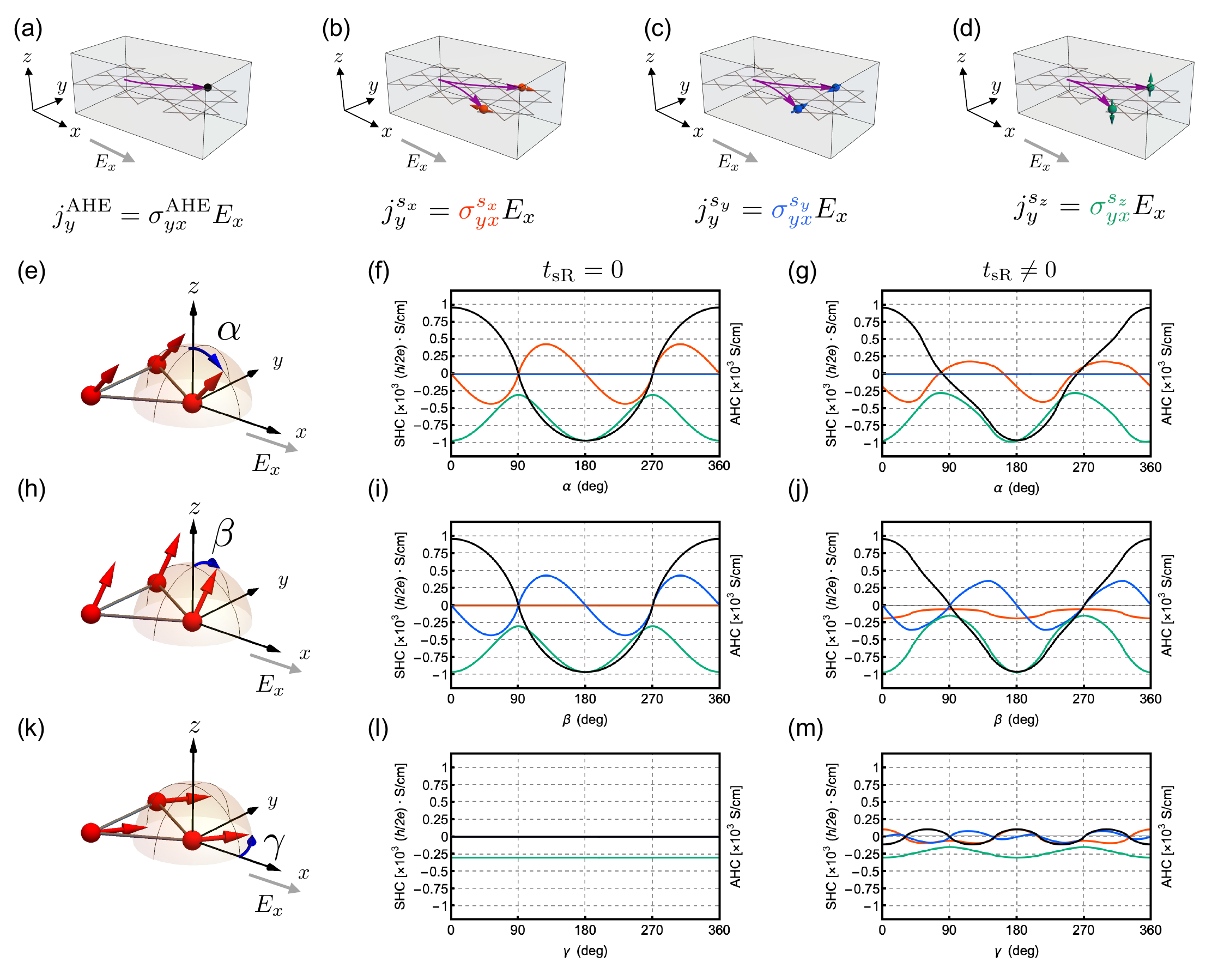}
\caption{(Color online)~
Schematic figures of in-plane (a)~anomalous Hall effect and (b-d)~spin Hall effect induced by electric field $E_{x}$.
(a)~Anomalous Hall current flows to $y$-direction and is characterized by the Hall conductivity $\sigma^{\rm AHE}_{yx}$.
Spin Hall currents with spin angular momentum
(b)~$s_x$,
(c)~$s_y$, and 
(d)~$s_z$,
flow to $y$ direction and are characterized by the Hall conductivities $\sigma^{s_x}_{yx}$, $\sigma^{s_y}_{yx}$, and $\sigma^{s_z}_{yx}$, respectively.
$\sigma^{\rm AHC}_{yx}$ (black), $\sigma^{s_x}_{yx}$ (red), $\sigma^{s_y}_{yx}$ (blue), and $\sigma^{s_z}_{yx}$ (green) (f) without and (g) with staggered Rashba SOC $t_{\rm sR}$, as a function of tilting angle (e) $\alpha$~(in $x$-$z$ plane).
(i)(j)~The same for tilting angle (h)~$\beta$~(in $y$-$z$ plane).
(l)(m)~The same for tilting angle (k)~$\gamma$~(in $x$-$y$ plane).
}
\label{fig:ip}
\end{figure*}

Next, we explain the paramagnetic Dirac semimetal state and the SHE~\cite{Lau2022}.
In Fig.~\ref{fig:weyl_dirac}(e), the energy spectrum of the paramagnetic Dirac state is schematically shown.
Figure~\ref{fig:weyl_dirac}(f) shows the DOS as a function of the energy, using the parameter ${\bm m}=(0,0,0)$. 
We emphasize that the energy of the Dirac points is located on $E\sim 0.1$~eV, deviating from $E\_{F}$.
The Dirac semimetal state the SHE occurs, as schematically shown in Fig.~\ref{fig:weyl_dirac}(g).
We focus on the spin current driven by an electric field pointing in the $x$-direction.
The SHCs can be calculated as
\begin{align} \nonumber
 \sigma^{s_\mu}_{\nu x}=
  & \hbar \sum_{n\neq m} {\rm Im} \int_{\rm BZ}\frac{d^3k}{(2\pi)^3} 
    \frac{f(E_{n{\bm k}})-f(E_{m {\bm k}})}{(E_{n{\bm k}}-E_{m{\bm k}})^2} \\ 
  &\times
  \Braket{n\bm{k}|\hat{j}^{\mu}_{\nu}|m\bm{k}}\Braket{m \bm{k}|(-e\hat{v}_x)|n\bm{k}} 
 \label{eqn:kubo_shc_ip},
\end{align}
where $j^{\mu}_{\nu}$ is a spin-current operator with a
spin polarization $\mu$ and spatial direction $\nu$,
given by $j^{\mu}_{\nu}=\frac{1}{2}\{\frac{\hbar}{2}\sigma_\mu,v_{\nu} \}$~\cite{Sinova2015}.
We neglect an extrinsic contribution from impurities for simplicity.
Figure~\ref{fig:weyl_dirac}(h) shows the $\sigma^{s_\mu}_{yx}$ as a function of the energy.
Near the energy of the Dirac points, corresponding to the minimam of the DOS,
$\sigma^{s_\mu}_{yx}$ is maximized.
To obtain the maximized SHC originating from the Dirac points, 
an appropriate tuning of $E\_{F}$ is required~\cite{Lau2022}.

\section{In-plane spin/anomalous Hall effect}

\begin{figure*}[t]
\includegraphics[width=0.9\hsize]{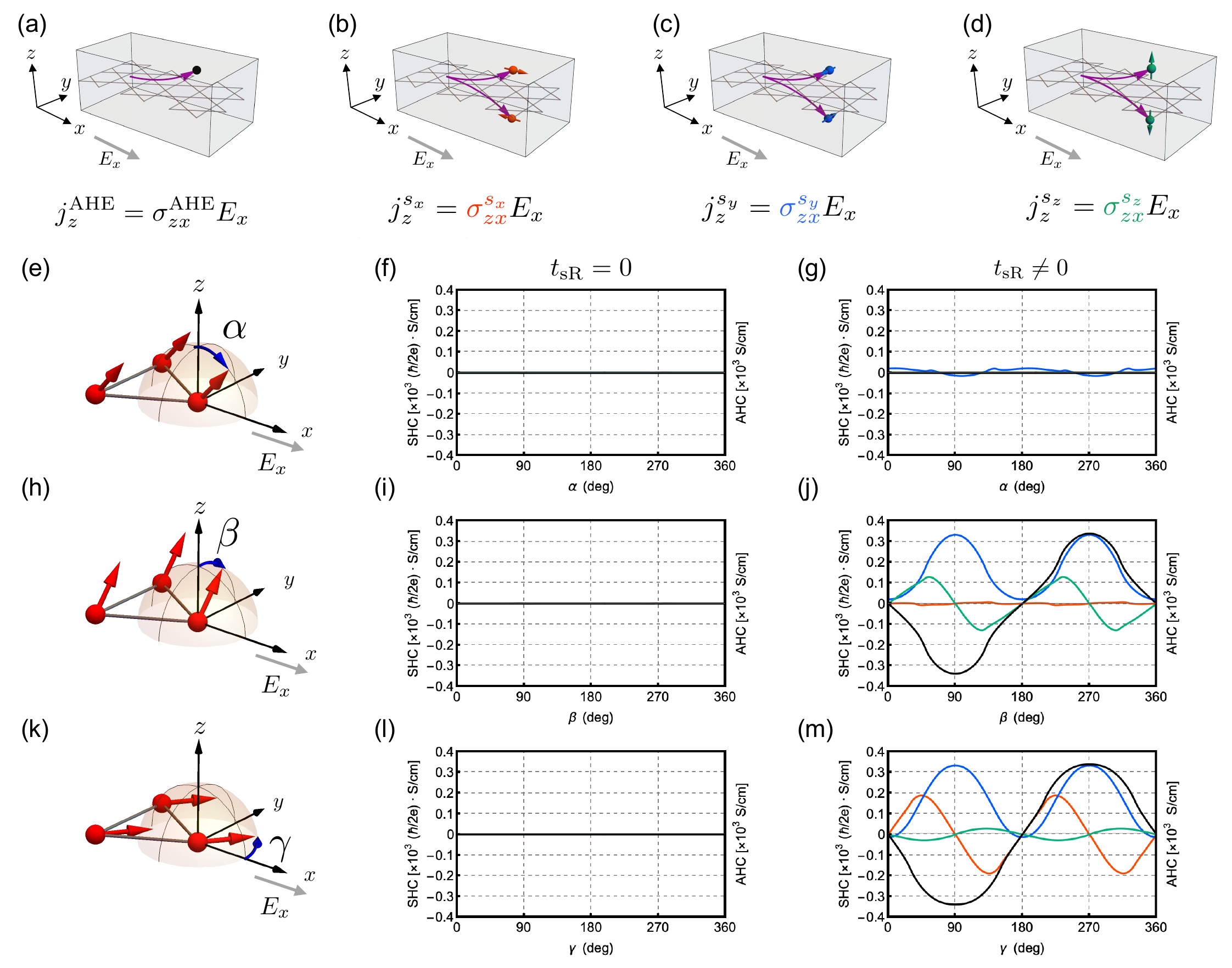}
\caption{(Color online)~
Schematic figures of out-of plane (a)~anomalous Hall effect and (b-d)~spin Hall effect induced by appleid electric field $E_{x}$.
(a)~Anomalous Hall current flows to $z$-direction and is characterized by the Hall conductivity $\sigma^{\rm AHE}_{zx}$.
Spin Hall currents with spin angular momentum
(b)~$s_x$,
(c)~$s_y$, and 
(d)~$s_z$,
flow to $z$-direction and are characterized by the spin Hall conductivities $\sigma^{s_x}_{zx}$, $\sigma^{s_y}_{zx}$, and $\sigma^{s_z}_{zx}$, respectively.
$\sigma^{\rm AHC}_{zx}$ (black), $\sigma^{s_x}_{zx}$ (red), $\sigma^{s_y}_{zx}$ (blue), and $\sigma^{s_z}_{zx}$ (green) (f) without and (g) with staggered Rashba SOC $t_{\rm sR}$, as a function of tilting angle (e) $\alpha$~(in $x$-$z$ plane).
(i)(j)~The same for different tilting angle (h)~$\beta$~(in $y$-$z$ plane).
(l)(m)~The same for different tilting angle (k)~$\gamma$~(in $x$-$y$ plane).
}
\label{fig:oop}
\end{figure*}

In the previous section, we reviewed that our model describes the SHE and AHE, 
originating from the non-magnetic Dirac state and ferromagnetic Weyl state, respectively.
The main purpose of this study is to clarify the relation between the SHE~(AHE) and the direction of the magnetic moment.
In this section, we study the in-plane SHE~(AHE), where the spin~(anomalous) Hall current $j^{s_{\mu}}_y$~($j^{\rm AHE}_y$) is induced by an electric field $E_x$, as schematically shown in Figs.~\ref{fig:ip}(a)-\ref{fig:ip}(d).
The in-plane SHCs and AHC are described as $\sigma^{s_\mu}_{yx}$ and $\sigma^{\rm AHE}_{yx}$, respectively.
Then, we introduce the magnetic moments with different directions, 
that are characterized by the parameter ${\bm m}_i$ in the exchange coupling term Eq.~(\ref{eqn:exc}).
We use uniform magnetization with three tilting angles in
(e)~$x$-$z$ plane $\alpha$,
(h)~$y$-$z$ plane $\beta$, and
(k)~$x$-$y$ plane $\gamma$,
as shown in Fig.~\ref{fig:ip}.
These magnetic configurations are given by,
(e) ${\bm m}_{A}={\bm m}_{B}={\bm m}_{C}=m(\sin\alpha,0,\cos\alpha)$,
(h) ${\bm m}_{A}={\bm m}_{B}={\bm m}_{C}=m(0,\sin\beta,\cos\beta)$,
(k) ${\bm m}_{A}={\bm m}_{B}={\bm m}_{C}=m(\cos\gamma,\sin\gamma,0)$,
respectively.
In \CSS, the magnetic moments may be tilted in the presence of the exchange bias effect~\cite{Seki2023} or the external magnetic field~\cite{Moghaddam2022}.
In all the following calculations, $E\_{F}$ is computed at each angle with the constant electron number~($n\_{e}=3$),
corresponding to the non-doped \CSS~\cite{Ozawa2019,Ozawa2022}.

First, we study the SHCs~(AHC) for the changes in the magnetization angle $\alpha$ shown in Fig.~\ref{fig:ip}(e).
In Fig.~\ref{fig:ip}(f), $\sigma^{s_\mu}_{yx}$ and $\sigma^{\rm AHE}_{yx}$ are computed as a function of $\alpha$
by considering only the Kane-Mele type SOC $t\_{KM}$.
When $\alpha=90^{\circ}$, 
the magnetization is parallel to the applied electric field.
Red, blue, green, and black lines indicate 
$\sigma^{s_x}_{yx}$,
$\sigma^{s_y}_{yx}$,
$\sigma^{s_z}_{yx}$, and
$\sigma^{\rm AHE}_{yx}$,
respectively.
The sign of the AHC $\sigma^{\rm AHE}_{yx}$~(black line) changes when the magnetization flips~\cite{Ozawa2019}. 
One finds that 
when $\alpha=0^{\circ}$ and $\alpha=180^{\circ}$,
$|\sigma^{\rm AHE}_{yx}|\sim|\sigma^{s_z}_{yx}|$~(green line) holds
except for the unit $\frac{\hbar}{2e}$
because the spin Hall current is equal to the spin polarized anomalous Hall current.
$\sigma^{s_x}_{yx}$~(red line) can be finite and has a periodicity of $180^{\circ}$.
On the other hand, $\sigma^{s_y}_{yx}$~(blue line) vanishes.
In Fig.~\ref{fig:ip}(g), the staggered-Rashba SOC $t\_{sR}$ is considered in addition to $t\_{KM}$.
We note $\sigma^{\rm AHE}_{yx}$ can be finite even without the out-of-plane component of the magnetization at $\alpha=90^{\circ}$ and $270^{\circ}$.
The periodicities of the Hall conductivities remain unchanged even with $t\_{sR}$.

Next, we study the SHCs~(AHC) for the changes in the magnetization angle $\beta$, as shown in Fig.~\ref{fig:ip}(h).
In Fig.~\ref{fig:ip}(i), $\sigma^{s_\mu}_{yx}$ and $\sigma^{\rm AHE}_{yx}$ are computed as a function of $\beta$ by considering only $t\_{KM}$.
When $\beta=90^{\circ}$, the magnetization is parallel to the spin Hall current.
The behaviors of $\sigma^{\rm AHE}_{yx}$~(black line) and $\sigma^{s_z}_{yx}$~(green line) are equivalent to those for the angle $\alpha$~[Fig.~\ref{fig:ip}(f)].
We find $\sigma^{s_y}_{yx}$~(blue line) can be finite, 
whereas $\sigma^{s_x}_{yx}$~(red line) vanishes.
In Fig.~\ref{fig:ip}(j),
$t\_{sR}$ is considered in addition to $t\_{KM}$.
In contrast to the case with $\alpha$, 
all the components of SHCs 
become finite. 
$\sigma^{\rm AHE}_{yx}$~(black line) vanishes at $\beta=90^{\circ}$ and $270^{\circ}$, 
while $\sigma^{\rm AHE}_{yx}$ can be finite at $\alpha=90^{\circ}$ and $270^{\circ}$.

Then, we study the SHCs~(AHC) for the changes in the magnetization angle $\gamma$ shown in Fig.~\ref{fig:ip}(k).
When $\gamma=0^{\circ}$, the magnetization is parallel to the applied electric field.
As shown in Fig.~\ref{fig:ip}(l), all the conductivities are independent of $\gamma$ 
by considering only $t\_{KM}$.
We note that only $\sigma^{s_z}_{yx}$~(green line) is finite.
In Fig.~\ref{fig:ip}(m),
$t\_{sR}$ is considered in addition to $t\_{KM}$.
All of the conductivities can be finite and show angular dependences.

\section{Out-of-plane spin/anomalous Hall effect}

In \CSS, it is experimentally favoreble to study the spin current flowing to the stacking~($z$-) direction~\cite{Lau2022,Seki2023}.
Motivated by such experiments,
we study the out-of-plane SHE~(AHE), 
where the spin~(anomalous) Hall current $j^{s_{\mu}}_z$~($j^{\rm AHE}_z$) 
is induced by an electric field $E_x$, as shown in Figs.~\ref{fig:oop}(a)-\ref{fig:oop}(d).
We find the characteristic angular dependences and enhancements of the Hall conductivities.
The out-of-plane SHCs and AHC are described as $\sigma^{s_\mu}_{zx}$ and $\sigma^{\rm AHE}_{zx}$, respectively.
We again consider the changes in the angles of the magnetization $\alpha$, $\beta$, and $\gamma$~[see Figs.~\ref{fig:oop}(e), \ref{fig:oop}(h), and \ref{fig:oop}(k)].
Here, red, blue, green, and black lines indicate 
$\sigma^{s_x}_{zx}$,
$\sigma^{s_y}_{zx}$,
$\sigma^{s_z}_{zx}$, and
$\sigma^{\rm AHE}_{zx}$,
respectively.
Figure~\ref{fig:oop} shows the out-of-plane Hall conductivities as a function of the angles as in Fig.~\ref{fig:ip}.
We notice that all the Hall conductivities vanish irrespective of the angles when only $t_{\rm KM}$ is finite [see Figs.~\ref{fig:oop}(f), \ref{fig:oop}(i), and \ref{fig:oop}(l)].
\begin{figure}[t]
\centering
\includegraphics[width=1\hsize]{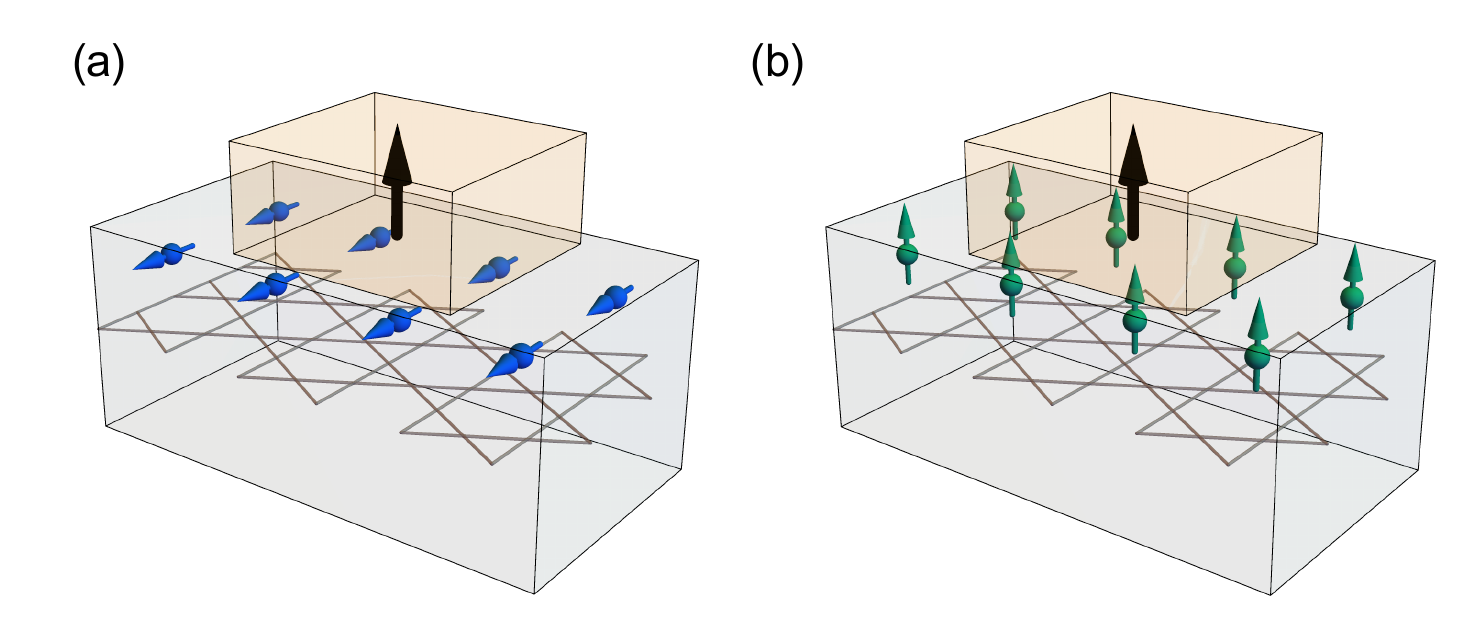}
\caption{(Color online)~Possible configurations of spin accumulation induced by out-of-plane spin Hall effect. 
(a)~$y$- and (b) $z$-component.
}
\label{fig:exp}
\end{figure}
Then, we consider $t\_{sR}$ in addition to $t\_{KM}$.
In Fig.~\ref{fig:oop}(g), for the angle $\alpha$, $\sigma^{s_y}_{zx}$~(blue line) can be finite, and others vanish.
In Figs.~\ref{fig:oop}(j) and \ref{fig:oop}(m), for the angles $\beta$ and $\gamma$, all of the Hall conductivities can be finite.
$\sigma^{s_y}_{zx}$~(blue line) and $|\sigma^{\rm AHE}_{zx}|$~(black line) are maximized at $\beta=\gamma=90^{\circ}$ and $270^{\circ}$.
At those angles, we recall the magnetization is perpendicular to the applied electric field and induced Hall currents.
This enhancement of the out-of-plane AHC $\sigma^{\rm AHE}_{zx}$ is consistent with that obtained by first-principles calculations~\cite{Ghimire2019}.
In Fig.~\ref{fig:oop}(j), for the angle $\beta$, $\sigma^{s_x}_{zx}$~(red line) is negligible.
On the other hand, in Fig.~\ref{fig:oop}(m), for the angle $\gamma$,
$\sigma^{s_x}_{zx}$~(red line) and $\sigma^{s_y}_{zx}$~(blue line) are comparable.
In particular, in both angles $\beta$ and $\gamma$, $|\sigma^{s_z}_{zx}|$~(green line) is enhanced around $60^{\circ}$, $120^{\circ}$, $240^{\circ}$, and $300^{\circ}$ while it vanishes at $0^{\circ}$, $90^{\circ}$, $180^{\circ}$, and $270^{\circ}$.
We emphasize that the distinct angular dependences of the Hall conductivities are originating from the presence of the staggered-Rashba type SOC.

Before concluding this paper, 
we discuss the possibility of this system as a functional spin current generator.
In the previous paragraph,
we found the enhancement of the SHCs by tilting the magnetization.
In particular, we showed the enhancement of $\sigma^{s_y}_{zx}$~(blue line) and $\sigma^{s_z}_{zx}$~(green line) as shown in Fig.~\ref{fig:oop}(j). 
This implies that, the (a)~$y$- and (b)~$z$-components of the spin accumulations are induced at the interface as shown in Fig.~\ref{fig:exp}.
Let us consider the geometry where the ferromagnet with perpendicular magnetization is attached to \CSS.
The spin accumulations induced by the SHE might be useful for the perpendicular magnetization switching, 
which is highly demanded from point of view of the device integration~\cite{fukami2016}. 
Figure~\ref{fig:exp} schematically shows the spin accumulations 
(a)~$\Braket{s_y}$ induced by $j^{s_y}_z=\sigma^{s_y}_{zx}E_x$ and 
(b)~$\Braket{s_z}$ induced by $j^{s_z}_z=\sigma^{s_z}_{zx}E_x$.
The spin accumulation $\Braket{s_y}$ gives a strong torque to the perpendicular magnetization. 
Its strength is as large as that in Pt at room temperature~\cite{Guo2008}.
The out-of-plane spin accumulation $\Braket{s_z}$ appears when the angle of magnetization of \CSS is tilted from the $z$-axis.
This $z$-component of the spin accumulation is useful for the deterministic spin torque~\cite{fukami2016,Hu2022},
which stablizes the magnetization switching as demanded in spintronics field.

\section{Conclusion}
In this paper we theoretically studied the SHE in an effective model of magnetic Weyl semimetal \CSS.
We showed the drastic changes of the SHCs depending on the direction of the magnetic moment.
Especially, for the out-of-plane SHE, the enhancements of the SHCs were found by considering the staggered-Rashba type SOC.
Our finding may help us design a functional spin current generator for the perpendicular magnetization switching based on the magnetic Weyl semimetal.

\acknowledgments
The authors would like to appreciate 
Y.~Araki,
K.~Fujiwara,
Y.~Kato,
Y.-C.~Lau,
Y.~Motome,
K.~Nakazawa,
T.~Seki, 
A.~Tsukazaki, and
Y.~Yahagi
for valuable discussions.
This work was supported by
JST CREST, Grant No.~JPMJCR18T2
and by
JSPS KAKENHI, Grant Nos.~
JP20H01830
and
JP22K03446.
A.~O.~was supported by
GP-Spin at Tohoku University 
and by
JST SPRING, Grant No.~JPMJSP2114.

\bibliography{main}

\end{document}